\documentclass[12pt,a4paper]{elsart}

\usepackage{amsmath,amssymb,array}
\usepackage{graphicx}
\usepackage{time}
\def\Today{\number\year\space
 \ifcase\month\or January\or February\or March\or April\or May\or June\or
    July\or August\or September\or October\or November\or December\fi
         \space\number\day
}

\usepackage[colorlinks,anchorcolor=blue,citecolor=blue,urlcolor=blue,pdfauthor=Keith Briggs and Christian Beck,pdfmenubar=false,pdffitwindow=true,pdfwindowui=false,hypertexnames=false]{hyperref}

\begin{document}
\begin{frontmatter}

\title{Modelling train delays with $q$-exponential functions}

\author{Keith Briggs}
\address{Complexity Research Group, BT,
Adastral Park, IP5 3RE, UK}

\author{Christian Beck}
\address{School of Mathematical Sciences, Queen Mary, University of
London, Mile End Road, London E1 4NS, UK}

\begin{abstract}
We demonstrate that the distribution of train delays on the
British railway network is accurately described by $q$-exponential
functions. We explain this by constructing an underlying superstatistical model.
\end{abstract}


\end{frontmatter}

\section{Introduction}

Complex systems in physics, engineering, biology, economics, and
finance, are often characterized by the occurence of fat-tailed
probability distributions. In many cases there is an asymptotic
decay with a power-law. For these types of systems more general
versions of statistical mechanics have been developed, in which
power laws are effectively derived from maximization principles
of more general entropy functions, subject to suitable
constraints \cite{tsa1,tsa2,tsa3,abe}. Typical distributions that
occur in this context are of the $q$-exponential form. The
$q$-exponential is defined as $e_q(x):=(1+(q-1)x)^{1/(q-1)}$,
where $q$ is a real parameter, the entropic index. It has become
common to call the corresponding statistics `$q$-statistics'.

A possible dynamical reason for $q$-statistics is a so-called
superstatistics \cite{cohen-beck}. For superstatistical complex
systems one has a superposition of ordinary local equilibrium
statistical mechanics in local spatial cells, but there is a
suitable intensive parameter $\beta$ of the complex system that
fluctuates on a relatively large spatio-temporal scale. This
intensive parameter may be the inverse temperature, or the
amplitude of noise in the system, or the energy dissipation in
turbulent flows, or an environmental parameter, or simply a local
variance parameter extracted from a suitable time series
generated by the complex system \cite{BCS}.
The superstatistics approach has been the subject of various
recent papers
\cite{beck-su,touchette,souza,chavanis,plastino,rajagopal} and it
has been applied to a variety of complex driven systems, such as
Lagrangian\cite{beck03,reynolds} and Eulerian
turbulence\cite{beck-physica-d,BCS}, defect
turbulence\cite{daniels}, cosmic ray statistics\cite{cosmic},
solar flares \cite{maya}, environmental turbulence
\cite{rapisarda}, hydroclimatic fluctuations \cite{hydro}, random
networks \cite{abe-thurner}, random matrix theory \cite{RMT} and
econophysics \cite{ausloos}.

If the parameter $\beta$ is distributed according to a particular
probability distribution, the $\chi^2$-distribution, then the
corresponding superstatistics, obtained by integrating over all
$\beta$, is given by $q$-statistics \cite{tsa1,tsa2,tsa3,abe},
which means that there are $q$-exponentials and asymptotic power
laws. For other distributions of the intensive parameter $\beta$,
one ends up with more general asymptotic decays \cite{touchette}.

In this paper we intend to analyse yet another complex system where
$q$-statistics seem to play an important role, and where a
superstatistical model makes sense. We have analysed in detail
the probability distributions of delays occuring on the British
rail network. The advent of real-time train information on the
internet for the British network
(http://www.nationalrail.co.uk/\- ldb/livedepartures.asp) 
has
made it possible to gather a large amount of data and therefore
to study the distribution of delays. Information on such delays
is very valuable to the traveller. Published information is
limited to a single point of the distribution - for example, the
fraction of trains that arrive with 5 minutes of their scheduled
time. Travellers thus have no information about whether the
distribution has a long tail, or even about the mean delay.   We
find that the delays are well modelled by a $q$-exponential
function, allowing a characterization of the distribution by two
parameters, $q$ and $b$. We will relate our observations to a
superstatistical model of train delays.

This paper is organized as follows:
first, we describe our data and the methods used for the
analysis. We then present our fitting results. In particular, we
will demonstrate that $q$-exponentials provide a good fit of the
train delay distributions, and we will show which parameters
$(q,b)$ are relevant for the various British rail network lines.
In the final section, we will discuss a superstatistical model
for train delays.

\section{The data}

We collected data on departure times for 23 major stations for the period
September 2005 to October 2006, by software which downloads the real-time
information webpage every minute for each station.   As each train actually
departs, the most recent delay value is saved to a database.   The database now
contains over two million train departures; for a busy station such as
Manchester Piccadilly over 200,000 departures are recorded.

\section{The model and parameter estimation}

\newcommand\BETA{b}

Preliminary investigation led us to believe that the model
\begin{equation}
e_{q,\BETA,c}(t)=c(1+\BETA(q-1)t)^{1/(1-q)}
\end{equation}
would fit well; here $t$ is the delay, $0\!<q\!<2$ and $\BETA\!>\!0$ are shape parameters, and $c$ is a normalization parameter.   We have
$e_{q,\BETA,c}(t)=c(1-\BETA t)+O(t^2)$ as $t\rightarrow 0$ and
$\log(e_{q,\BETA,c}(t))/\log(t)\rightarrow 1/(1-q)$ as  $t\rightarrow\infty$.
These limiting forms allow an initial estimate of the parameters; an accurate
estimate is then obtained by nonlinear least-squares.   We also have 
\begin{equation}
\lim_{q\rightarrow 1}e_{q,\BETA,c}(t)=c\exp(-\BETA t),
\end{equation}
so that $q$ measures the deviation from an exponential distribution.   An
estimated $q$ larger than unity indicates a long-tailed distribution.
 
We did not include the zero-delay value in the fitted models.   Typically 80\%
of trains record $t=0$, indicating a delay of one minute or less (the
resolution of the data).  Thus, our model represents the conditional
probability distribution of the delay, given that the train is delayed one
minute or more.

In order to provide meaningful parameter confidence intervals, we weighted the
data as follows.  Since our data is in the form of a histogram, the
distribution of the height $c_i$ of the bar representing the count of trains
with delay $i$ will be binomial. In fact, it is of course very close to Gaussian
whenever $c_i$ is large enough, which is the case nearly always.   The
normalized height $f_i=c_i/n$ (where $n$ is the total number of trains) will
therefore have standard deviation $\sigma_i=(n f_i(1-f_i))^{1/2}/n\approx
c_i^{1/2}/n$.   We used these values as weights in the nonlinear least squares
procedure, and hence computed parameter confidence intervals by standard
methods, namely from the estimated parameter covariance matrix.  We find that typically $q$ and $b$ have a correlation coefficient of about $-0.5$; thus, the very small confidence intervals quoted in the figure captions for $b$ are not particularly useful; $b$ typically acquires a larger uncertainty via its correlation with $q$.

\section{Results}

\begin{figure}[ht]
\begin{center}
\includegraphics[width=0.8\hsize]{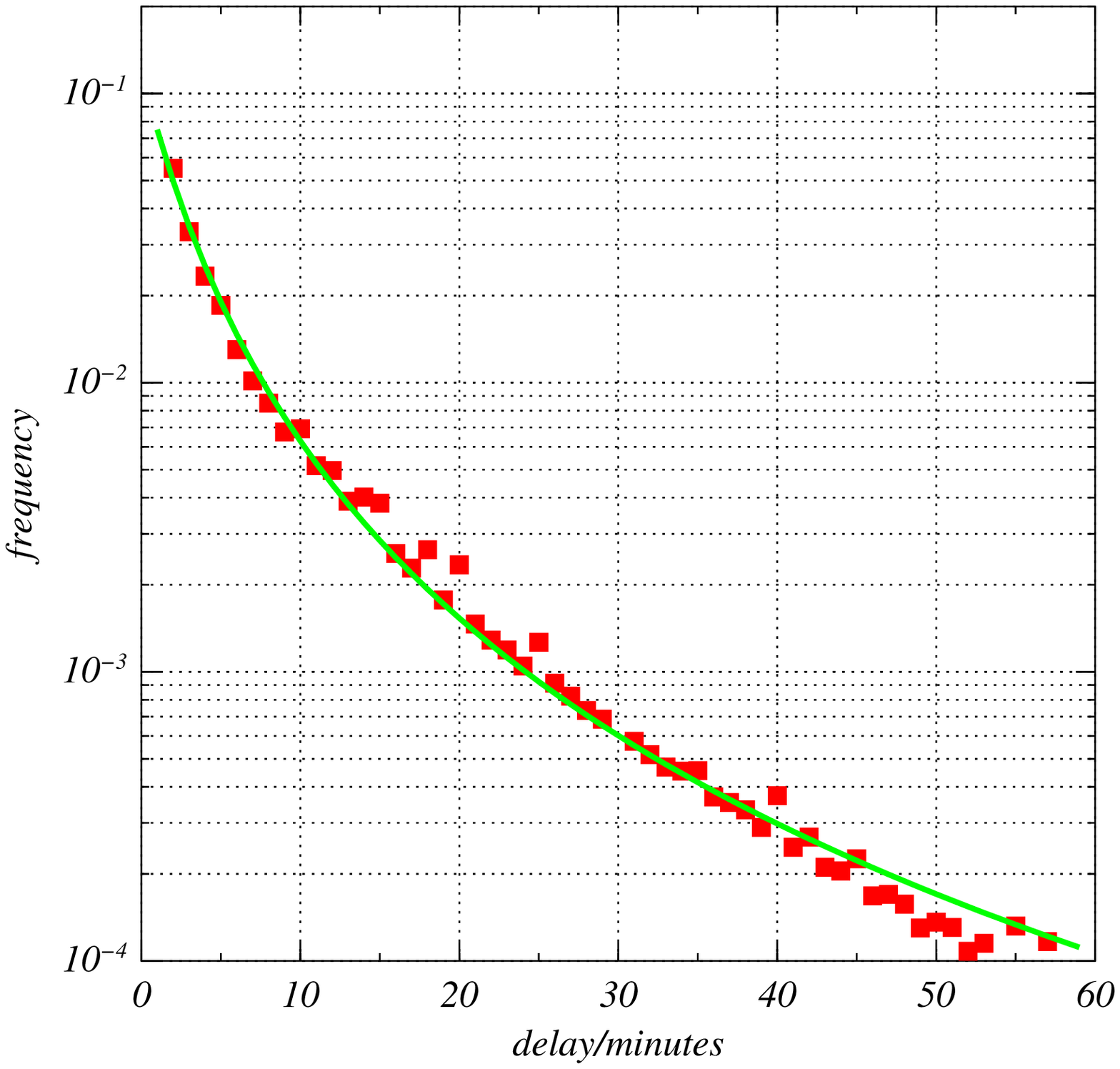}
\caption{
  All train data and best-fit $q$-exponential:
  $q=1.355 \pm 8.8\times 10^{-5}$,
  $b=0.524 \pm 2.5\times 10^{-8}$.
}
\label{_}
\end{center}
\end{figure}

\begin{figure}[ht]
\begin{center}
\includegraphics[width=0.8\hsize]{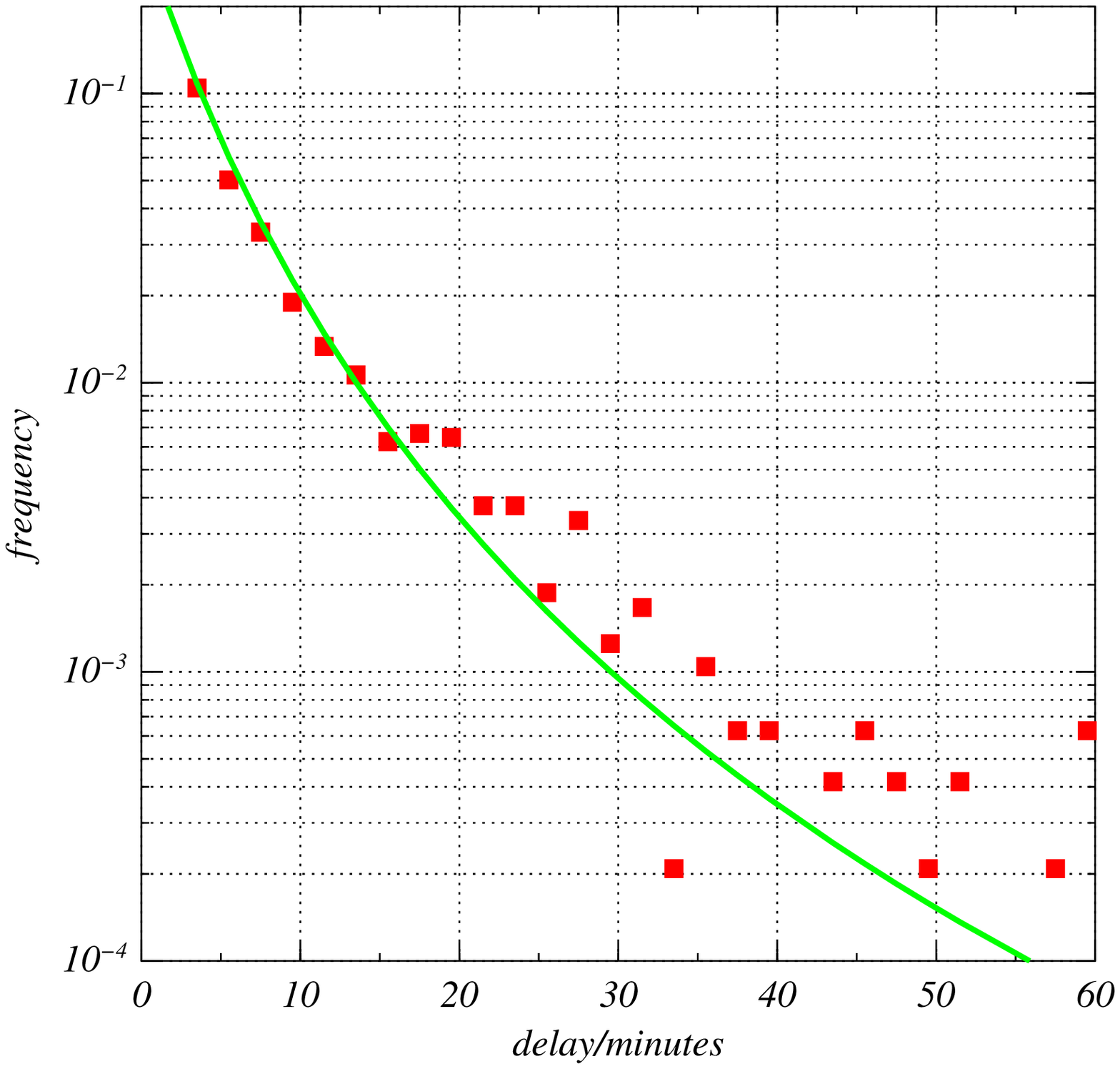}
\caption{
  Bath Spa to London Paddington (showing typical fluctuations in the tail when data is sparse):
  $q=1.215 \pm 0.015$,
  $b=0.405 \pm 2.8\times 10^{-6}$.
}
\label{BTH_PAD}
\end{center}
\end{figure}

\begin{figure}[ht]
\begin{center}
\includegraphics[width=0.8\hsize]{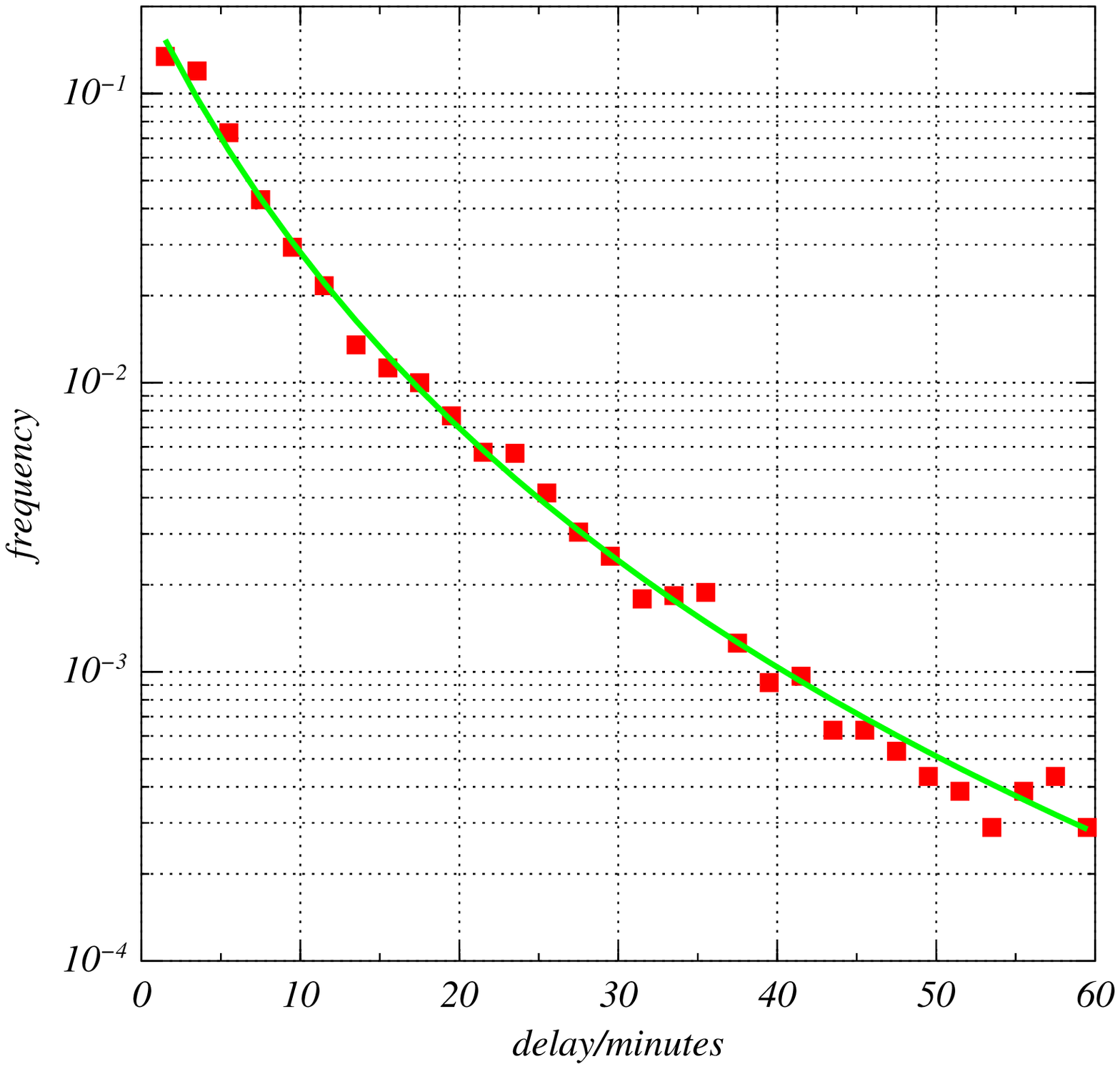}
\caption{
  Swindon to London Paddington:
  $q=1.230 \pm 0.0086$,
  $b=0.266 \pm 3.1\times 10^{-6}$.
}
\label{SWI_PAD}
\end{center}
\end{figure}

\begin{figure}[ht]
\begin{center}
\includegraphics[width=0.8\hsize]{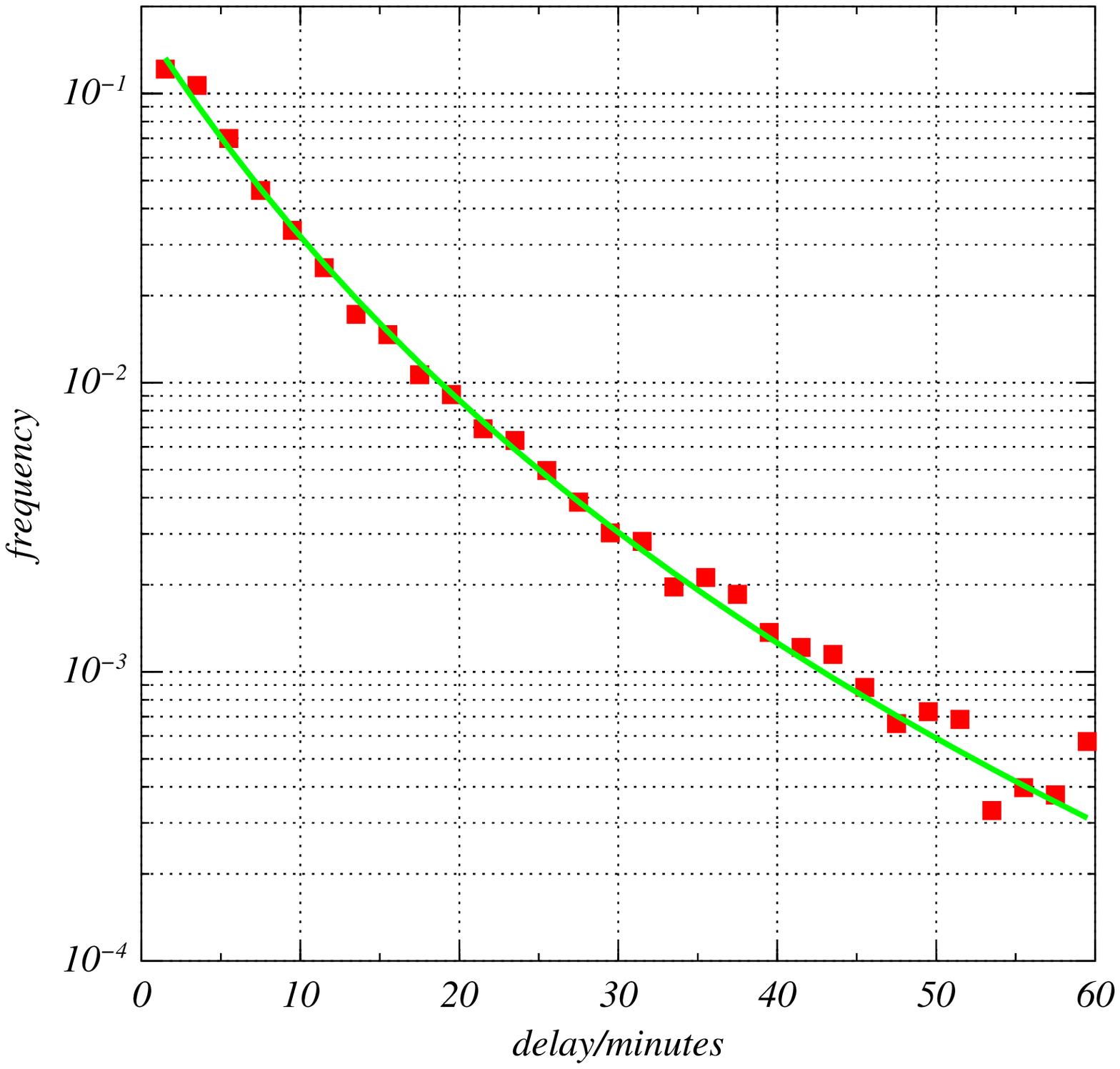}
\caption{
  Reading to London Paddington:
  $q=1.183 \pm 0.0063$,
  $b=0.202 \pm 2.7\times 10^{-6}$.
}
\label{RDG_PAD}
\end{center}
\end{figure}

We first fitted the model to all data, obtaining the fit shown in
Figure~\ref{_}.   This corresponds to a `universality' assumption - if all
routes had the same distribution of delays, the parameter values $q=1.355,
b=0.524$ would be the relevant ones.   We may thus compare the parameters
for specific routes with these.  Typical fits for three such routes are shown in
 Fig.~\ref{BTH_PAD}, 
Fig.~\ref{SWI_PAD}, and Fig.~\ref{RDG_PAD}.

Delays typically build up over a train's journey, and are very unlikely at the
initial departure station.   Thus, we choose to study delays at intermediate
stations.  At such stations, a delayed departure almost certainly means the
arrival was delayed.  



\section{Superstatistical model}

We start with a very simple model for the local departure
statistics of trains. The waiting time distribution until
departure takes place is simply given by that of a Poisson process
\cite{vKa}
\begin{equation}
P(t|\beta)=\beta e^{-\beta t}.
\end{equation}
Here $t$ is the time delay from the scheduled departure time, and
$\beta$ is a positive parameter. The symbol $P(t|\beta)$ denotes
the conditional probability density to observe the delay $t$
provided the parameter $\beta$ has a certain given value.
Clearly, the above probability density is normalized. Large
values of $\beta$ mean that most trains depart very well in time,
whereas small $\beta$ describe a situation where long delays are
rather frequent.

The above simple exponential model becomes superstatistical by
making the parameter $\beta$ a fluctuating random variable as
well. These fluctuations describe large-scale temporal variations
of the British rail network environment. For example, during the
start of the holiday season, when there is many passengers,
we expect that $\beta$ is smaller than usual for a while,
resulting in frequent delays. Similarly, if there is a problem
with the track or if bad weather conditions exist,
we also expect smaller values of $\beta$ on
average. The value of $\beta$ is also be influenced by extreme
events such as derailments, industrial action, terror alerts, etc.

The observed long-term distribution of train delays is then a
mixture of exponential distributions where the parameter $\beta$
fluctuates. If $\beta$ is distributed with probability density
$f(\beta)$, and fluctuates on a large time scale, then one
obtains the marginal distributions of train delays as
\begin{equation}
p(t)=\int_0^\infty f(\beta) p(t|\beta) d\beta = \int_0^\infty f(\beta) \beta e^{-\beta t}.
\label{9}
\end{equation}
It is this marginal distribution that is actually recorded in our
data files.

Let us now construct a simple model for the distribution
$f(\beta)$. There may be $n$ different Gaussian random variables
$X_i, i=1,\ldots,n$, that influence the dynamics of the positive
random variable $\beta$ in an additive way \cite{prl}. We may thus assume as
a very simple model that
\begin{equation}
\beta =\sum_{i=1}^n X_i^2,
\end{equation}
where $\langle X_i \rangle =0$ and $\langle X_i^2 \rangle \not=0$.
In this case the probability density of $\beta$ is given by a
$\chi^2$-distribution with $n$ degrees of freedom:
\begin{equation}
f (\beta) = \frac{1}{\Gamma \left( \frac{n}{2} \right)} \left(
\frac{n}{2\beta_0}\right)^{\frac{n}{2}} \beta^{\frac{n}{2}-1}
\exp\left(-\frac{n\beta}{2\beta_0} \right). \label{fluc}
\end{equation}
The average  of $\beta$ is given by
\begin{equation}
\langle \beta \rangle =n\langle X_i^2\rangle=\int_0^\infty\beta
f(\beta) d\beta= \beta_0
\end{equation}
and the variance by
\begin{equation}
\langle \beta^2 \rangle -\beta_0^2=
\frac{2}{n} \beta_0^2.
\end{equation}

The integral (\ref{9}) is easily evaluated and one obtains
\begin{equation}
p(t) \sim {\left( 1+b(q-1)t\right)^{\frac{1}{1-q}}}
\end{equation}
where $q=1+{2}/(n+2)$ and $b= {2}\beta_0 /({2-q})$.
Our model generates $q$-exponential distributions of train delays
by a simple mechanism, namely a $\chi^2$-distributed parameter
$\beta$ of the local Poisson process.

\begin{figure}[ht]
\begin{center}
\includegraphics[width=0.8\hsize]{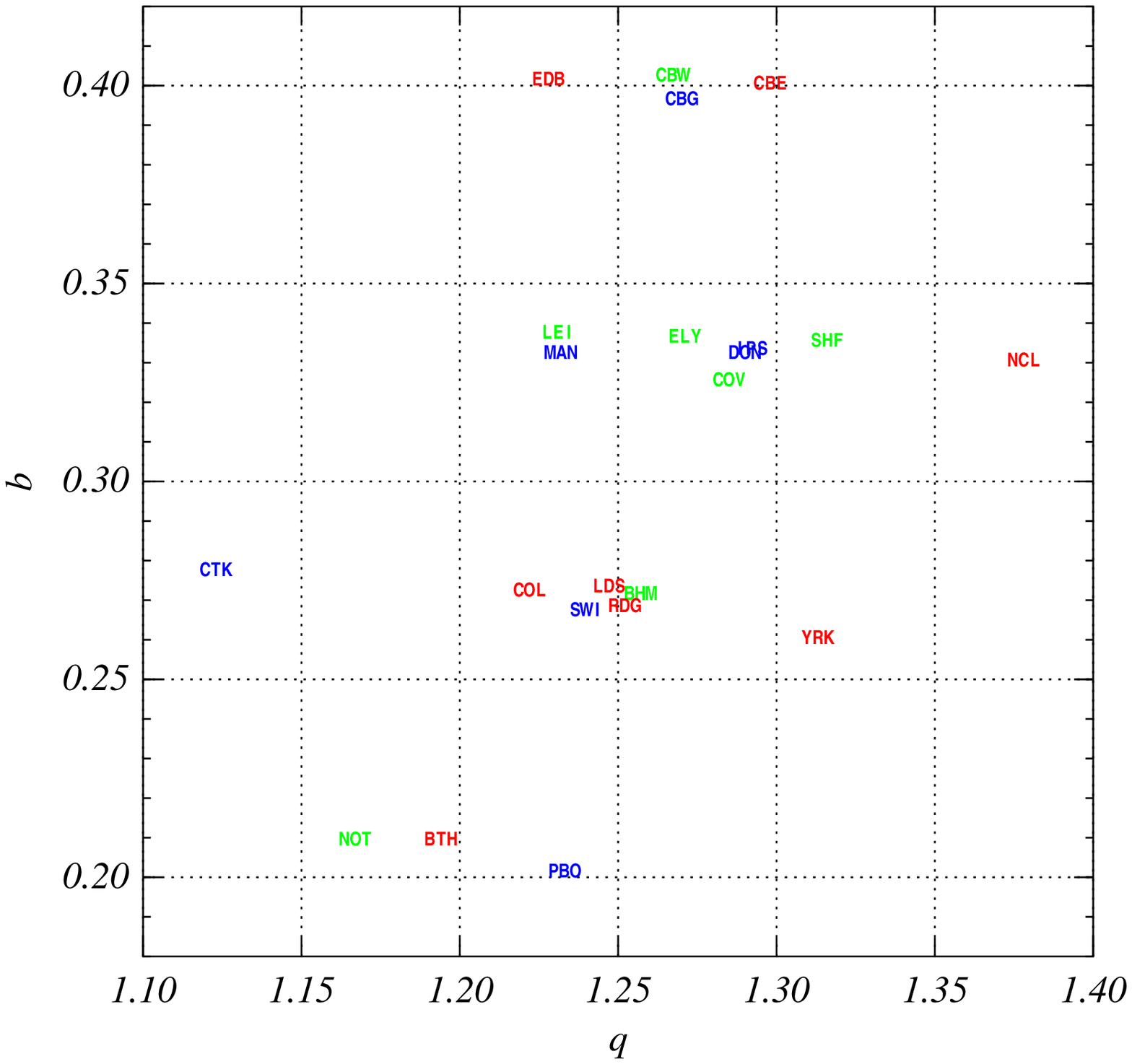}
\caption{The estimated parameters $q$ and $b$ for 23 stations.
}
\label{q_b_plot}
\end{center}
\end{figure}

\begin{table}[ht]
\begin{center}
\setlength{\extrarowheight}{-1.5mm}
\begin{tabular}{lccl}
\hline\strut
station & $q$ & $b$ & code\\
\hline
Bath Spa& 1.195 & 0.209 & BTH\strut\\
Birmingham & 1.257 & 0.271 & BHM\\
Cambridge & 1.270 & 0.396 & CBG\\
Canterbury East & 1.298 & 0.400 & CBE\\
Canterbury West & 1.267 & 0.402 & CBW\\
City Thameslink & 1.124 & 0.277 & CTK\\
Colchester & 1.222 & 0.272 & COL\\
Coventry & 1.291 & 0.330 & COV\\
Doncaster & 1.289 & 0.332 & DON\\
Edinburgh & 1.228 & 0.401 & EDB\\
Ely & 1.316 & 0.393 & ELY\\
Ipswich & 1.291 & 0.333 & IPS\\
Leeds & 1.247 & 0.273 & LDS\\
Leicester & 1.231 & 0.337 & LEI\\
Manchester Piccadilly & 1.231 & 0.332 & MAN\\
Newcastle & 1.378 & 0.330 & NCL\\
Nottingham & 1.166 & 0.209 & NOT\\
Oxford & 1.046 & 0.141 & OXF\\
Peterborough & 1.232 & 0.201 & PBO\\
Reading & 1.251 & 0.268 & RDG\\
Sheffield & 1.316 & 0.335 & SHF\\
Swindon & 1.226 & 0.253 & SWI\\
York & 1.311 & 0.259 & YRK
\end{tabular}
\caption{The estimated parameters $q$ and $b$ for 23 stations.}
\label{q_b_table}
\end{center}
\end{table}

Typical $q$-values obtained from our fits are in the region
$q=1.15 \dots 1.35$ (see Fig.~\ref{q_b_plot} and Table~\ref{q_b_table}). 
This means
\begin{equation}
n=\frac{2}{q-1}-2
\end{equation}
is in the region $4\ldots 11$. This means the number of degrees
of freedom influencing the value of $\beta$ is just of the order
we expected it to be: A few large-scale phenomena such as
weather, seasonal effects, passenger fluctuations, signal
failures, repairs of track, etc. seem to be relevant.

We can also estimate the average contribution of each degree of
freedom, from the fitted value of $b$. We obtain
\begin{equation}
\langle X_i^2 \rangle =\frac{\beta_0}{n}
=\frac{2-q}{n}b=\frac{1}{2}(q-1)b.
\end{equation}
If the above number is large for a given station, the local
station management seems to be doing a good job, since in this
case the local exponential decay of the delay times is as fast as
it can be. In general, it makes sense to compare stations with the
same $q$ (the same number of external degrees of freedom of the
network environment): The larger the value of $b$, the better the
performance of this station under the given environmental
conditions. Our analysis shows that two of the best performing busy stations
according to this criterion are Cambridge and Edinburgh.

\end{document}